\documentclass[11pt]{article}
\newcommand{\rL}{{\rm L}}

\def\ptoday{{\sl {\number\day \space de\space \ifcase\month 
\or janeiro\or fevereiro\or mar{\c c}o\or abril\or maio
\or junho\or julho\or agosto\or setembro\or outubro
\or novembro \or dezembro\fi\space de\space \number\year}}}    
\usepackage{amsmath}
\usepackage{amsfonts,amssymb,latexsym}
\usepackage{mathrsfs}
\usepackage{amscd}
\usepackage{amsthm}

\theoremstyle{remark}

\makeatletter
\@addtoreset{equation}{section}
\makeatother

\newcommand{\ddouble}{{\partial^{^{\kern-6pt \leftrightarrow}}}}

\newcommand{\dpad}[2]{{\displaystyle{\frac{\partial #1}{\partial #2}}}}

\newcommand{\equ}[1]{(\ref{#1})}

\newcommand{\p}{\psi}
\newcommand{\beq}{\begin{equation}}
\newcommand{\eqn}[1]{\label{#1}\end{equation}}
\newcommand{\ba}{\begin{array}}
\newcommand{\ea}{\end{array}}
\newcommand{\journal}[4]{{#1~}{\bf #2}~(19#3)~#4.}

\newcommand{\pr}{\journal {Phys. Rev.}}

\newcommand{\prl}{\journal {Phys. Rev. Lett.}}

\newcommand{\cmp}{\journal {Commun. Math. Phys.}}

\newcommand{\np}{\journal {Nucl. Phys.}}
\newcommand{\pl}{\journal {Phys. Lett.}}

\newcommand{\prep}{\journal {Phys. Rep.}}

\newcommand{\nc}{\journal {Nuovo Cimento}}

\newcommand{\annp}{\journal {Ann. Phys. (N.Y.)}}

\def\a{\alpha}
\def\b{\beta}
\def\d{\delta}
\def\e{\epsilon}

\def\vf{\varphi}
\def\g{\gamma}

\def\l{\lambda}
\def\m{\mu}
\def\n{\nu}

\def\s{\sigma}

\def\D{\Delta}

\def\G{\Gamma}

\def\S{\Sigma}

\def\inbar{\vrule height1.5ex width.4pt depth0pt}
\def\rlx{\relax\leavevmode}
\def\I{\leavevmode\hbox{\small1\kern-3.8pt\normalsize1}}
\def\openone{\leavevmode\hbox{\small1\kern-3.3pt\normalsize1}}
\def\Ione{\rlx{\rm 1\kern-2.7pt l}}
\def\Ik{\rlx{\rm I\kern-.18em k}}
\def\IC{\rlx\leavevmode
             \ifmmode\mathchoice
                    {\hbox{\kern.33em\inbar\kern-.3em{\rm C}}}
                    {\hbox{\kern.33em\inbar\kern-.3em{\rm C}}}
                    {\hbox{\kern.28em\sinbar\kern-.25em{\rm C}}}
                    {\hbox{\kern.25em\ssinbar\kern-.22em{\rm C}}}
             \else{\hbox{\kern.3em\inbar\kern-.3em{\rm C}}}\fi}
\def\IP{\rlx{\rm I\kern-.18em P}}
\def\IR{\rlx{\rm I\kern-.18em R}}
\def\IN{\rlx{\rm I\kern-.20em N}}

\def\llsymbol#1{\@llsymbol{\@nameuse{c@#1}}}
\def\@llsymbol#1{\ifcase#1\or {}\or {'}\or {''}\or {'''}\or
   {''''}\or {'''''}\or  \else\@ctrerr\fi\relax}
\newcommand{\ol}\overline
\newcommand{\ti}\tilde
\newcommand{\wt}\widetilde
\newcommand{\wh}\widehat
\newcommand{\bv}\breve
\newcommand{\dg}\dagger

\newcommand{\be}{\begin{equation}}
\newcommand{\ee}{\end{equation}}
\newcommand{\bl}{\begin{eqnarray}&}
\newcommand{\el}{&\end{eqnarray}}
\newcommand{\bq}{\begin{eqnarray}}
\newcommand{\eq}{\end{eqnarray}}

\newcommand{\pa}{\partial}

\def\sl#1{\rlap{\hbox{$\mskip 1 mu /$}}#1}

\def\ssl#1{\rlap{\hbox{$ {\scriptstyle /}$}}#1}

\begin{document}

\begin{center}
{{\LARGE \bf{Lorentz and CPT violation in QED revisited:\\[3mm] 
A missing analysis}}}

\vspace{7mm}

{\large Oswaldo M. Del Cima$^{\rm (a)}$\footnote{{{e-mail: wadodelcima@if.uff.br}}},
Jakson M. Fonseca$^{\rm (b)}$\footnote{{{e-mail: jakson.fonseca@ufv.br}}},\\ 
Daniel H. T. Franco$^{\rm (b)}$\footnote
{{{e-mail: daniel.franco@ufv.br}}} 
and Olivier Piguet$^{\rm (c)}$\footnote{{{e-mail: opiguet@pq.cnpq.br}}}}

\vspace{4mm}

$^{{\rm (a)}}$ {\it Universidade Federal Fluminense (UFF), P\'olo Universit\'ario de
                    Rio das Ostras\\
                    Rua Recife s/n - 28890-000 - Rio das Ostras - 
                    RJ - Brasil.}

$^{{\rm (b)}}$ {\it Universidade Federal de Vi\c cosa (UFV), Departamento de F\'\i sica\\
                    Avenida Peter Henry Rolfs s/n - 36570-000 - Vi\c cosa - 
                    MG - Brasil.}

$^{{\rm (c)}}$ {\it Universidade Federal do Esp\'\i rito Santo (UFES), Departamento de F\'\i sica\\
                    Campus Universit\'ario de Goiabeiras - 29060-900 - Vit\'oria - ES - Brasil.}

\end{center}

\begin{center}
{\bf Abstract}
\end{center}

{\small  
We investigate the breakdown of Lorentz symmetry in QED by a 
CPT violating interaction term consisting of the coupling of
an axial fermion current with a constant vector field $b$, 
in the framework of algebraic renormalization --  
a regularization-independent method. 
We show, to all orders in perturbation theory, that a 
CPT-odd and Lorentz violating Chern-Simons-like term, 
definitively, is not radiatively induced by the axial coupling 
of the fermions with the constant vector $b$.

\section{Introduction}
\hspace*{\parindent}
The quantum electrodynamics (QED) with violation of Lorentz and CPT have been studied
intensively in recent years. Among several issues, the possible generation of a
Chern-Simons-like term induced by radiative corrections arising from a CPT and Lorentz
violating term in the fermionic sector has been a recurrent theme in the literature.
We particularly mention the following works~\cite{Ref1}-\cite{Ref15} (and references
cited therein), where many controversies have emerged from the discussion whether this
Chern-Simons-like term could be generated by means of radiative corrections arising
from the axial coupling of charged fermions to a constant vector $b_\mu$ responsible
for the breakdown of Lorentz Symmetry.

In this paper, we reassess the discussion on the radiative generation of a
Chern-Simons-like term induced from quantum corrections in the extended QED.
Concerning to extended QED with a term which violates the
Lorentz and CPT symmetries, most of the papers were devoted to discuss
the gauge invariance of the model {\it only}, putting aside a more specific way how 
Lorentz invariance is broken. Here, we will discuss the latter point,
giving attention to the requirement that the breakdown of
Lorentz symmetry arising from the axial coupling of charged fermions
to a constant vector $b_\mu$ be {\it soft} in the sense of 
Symanzik~\cite{Ref16,Ref16',brs-break,piguet}, {\it i.e.}, has power-counting
dimension less than four or, equivalently, is negligeable in the deep Euclidean 
region of energy-momentum space. To the best of our knowledge this 
has not been investigated in details. In switching on the radiative corrections,
it is a non-trivial task to study the effects of such a symmetry breaking. 
In particular, one has to ask how the corresponding Ward identity that characterizes 
the  breaking behaves at the quantum level. Our aim is to show that, to the contrary
of the claims found in the literature, radiative corrections arising from the axial
coupling of charged fermions to a constant vector $b_\mu$ do not induce a Lorentz-
and CPT-violating Chern-Simons-like term in the QED action. 

\section{Extended QED in the Classical Approximation}
\label{sect2}
\subsection{The Classical Theory}
\hspace*{\parindent}
We start by considering an action for extended QED with a term which violates the
Lorentz and CPT symmetries in the matter sector only. In the tree approximation,
the classical action of extended QED with one Dirac spinor that we are considering
here is given by:
\begin{align}
\Sigma=\Sigma_{\rm S}+\Sigma_{\rm SB}+\Sigma_{\rm IR}+\Sigma_{\rm gf}\,\,,
\label{1}
\end{align}
where
\begin{align}
\Sigma_{\rm S}=
\int d^{4}x\,\,\Big\{i\bar{\psi}\gamma^{\mu}(\partial_{\mu}+ieA_{\mu})\psi -
m \bar{\psi}\psi-\frac{1}{4}F^{\mu\nu}F_{\mu\nu}\Big\}\,\,,
\end{align}
is the symmetric part of $\Sigma$ under gauge and Lorentz transformations.
 The term
\begin{align}
\Sigma_{\rm SB}=-\int d^{4}x\,\,b_{\mu}\bar{\psi}\gamma_{5}\gamma^{\mu}\psi\,\,, 
\label{SB}
\end{align}
is the symmetry-breaking part of $\Sigma$ that breaks the manifest Lorentz
covariance by the presence of a constant vector $b_{\mu}$ which selects a
preferential direction in Minkowski space-time, breaking its isotropy, as
well as it breaks CPT.\footnote{Greenberg proved that CPT invariance 
is necessary, but not sufficient, for Lorentz invariance \cite{greenberg}.}
\begin{align}
\Sigma_{\rm IR}=\int d^{4}x\,\,\frac{1}{2}\lambda^2A_\mu A^\mu\,\,,
\end{align}
is a mass term for the photon field,\footnote{As we shall see, the gauge
invariance properties are not spoiled by the photon mass: this is a peculiarity
of the Abelian case \cite{piguet-rouet}. This was studied in details for the QED
in Ref.\cite{low-schroer} using the BPHZ scheme.} introduced in order to avoid
infrared singularities and
\begin{align}
\Sigma_{\rm gf}=-\int d^{4}x\,\,\frac{1}{2\xi}\bigl(\partial_\mu A^\mu\bigr)^2\,\,,
\end{align}
is a gauge-fixing action.

\subsection{The Symmetries}
\subsubsection{Discrete Symmetries}
The discrete symmetries of the theory are the following ones.

\noindent {\bf Charge Conjugation C}:
Assuming the Dirac representation of the $\gamma$-matrices \cite{itzykson}, 
the charge conjugation transformations read: 
\begin{eqnarray}\label{9}
\psi &\stackrel{C}{\longrightarrow}& \psi^c=C\,{\bar{\psi}}^t\,\,,\nonumber \\
\bar{\psi} &\stackrel{C}{\longrightarrow}& \bar{\psi}^c=-\psi ^tC^{-1}\,\,,\nonumber \\
A_\mu &\stackrel{C}{\longrightarrow}& A_{\mu}^{c}=-A_{\mu}\,\,,\nonumber \\
C\gamma_{\mu}C & = & \gamma_{\mu}^{t}\,\,,\nonumber \\
C\gamma_{5}C & = & -\gamma_{5}^{t}=-\gamma_{5}\,\,.
\end{eqnarray}
where $C$ is the charge conjugation matrix, with $C^{2}=-1$. 
All terms in the action $\Sigma$ (\ref{1}) are invariant under
charge conjugation.

\noindent {\bf Parity P}:
\begin{eqnarray}\label{11}
x &\stackrel{P}{\longrightarrow}& (x^0,\,-\vec{x})\,\,,\nonumber \\
\psi &\stackrel{P}{\longrightarrow}& \gamma^{0}\psi\,\,,\nonumber \\
\bar{\psi} &\stackrel{P}{\longrightarrow}& \bar{\psi}\gamma^{0}\,\,,\nonumber \\
A_{\mu} &\stackrel{P}{\longrightarrow}& A^{\mu}\,\,.\nonumber \\
\end{eqnarray}
All terms of the action are invariant, excepted the Lorentz  breaking term
$\Sigma_{\rm SB}$ (\ref{SB}), which transforms under parity as
\begin{equation}\label{12}
\bar{\psi}b_{\mu}\gamma_{5}\gamma^{\mu}\psi
\stackrel{P}{\longrightarrow} \left\{
\begin{array}{ll}
-\bar{\psi}b_{0}\gamma_{5}\gamma^{0}\psi  \\
\bar{\psi}b_{i}\gamma_{5}\gamma^{i}\psi\,,\quad(i=1,2,3)\,.
\end{array}
\right.
\end{equation}

\noindent {\bf Time Reversal T}:
\begin{eqnarray}\label{13}
\psi &\stackrel{T}{\longrightarrow}& T\psi\,\,,\nonumber \\
\bar{\psi} &\stackrel{T}{\longrightarrow}& \bar{\psi}T\,\,,\nonumber \\
A_{\mu} &\stackrel{T}{\longrightarrow}& A_{\mu}\,\,,\nonumber \\
T\gamma^{\mu}T &=& \gamma_{\mu}^{T}=\gamma^{\mu*}\,\,,\nonumber \\
T\gamma_{5}T &=& \gamma_{5}\,\,.
\end{eqnarray}
Under time reversal transformation, the broken Lorentz term, $\Sigma_{\rm SB}$
(\ref{SB}), transforms as below:
\begin{equation}\label{14}
\bar{\psi}b_{\mu}\gamma_{5}\gamma^{\mu}\psi
\stackrel{T}{\longrightarrow} \left\{
\begin{array}{ll}
\bar{\psi}b_{0}\gamma_{5}\gamma^{0}\psi  \\
-\bar{\psi}b_{i}\gamma_{5}\gamma^{i}\psi\,,\quad(i=1,2,3)\,,
\end{array}
\right.
\end{equation}
which implies time reversal violation, 
whereas the other terms in the action $\Sigma$ (\ref{1}) remain invariant. 

Therefore, the action for extended QED, $\Sigma$ (\ref{1}), 
has  CPT symmetry broken by the Lorentz breaking
term, $\Sigma_{\rm SB}$ (\ref{SB}): 
\begin{equation}\label{15}
\bar{\psi}b_{\mu}\gamma_{5}\gamma^{\mu}\psi
\stackrel{CPT}{\longrightarrow}
-\bar{\psi}b_{\mu}\gamma_{5}\gamma^{\mu}\psi\,\,.
\end{equation}

\subsubsection{Continuous Symmetries: The Functional Identities}
\label{funct-ident}
\hspace*{\parindent}
The $U(1)$ gauge transformations are given by:
\begin{align}\label{gaugetransf}
\delta_{\rm g} A_{\mu}(x)=&\,\frac{1}{e}\,\partial_{\mu}\omega(x)\,\,,\nonumber\\[3mm]
\delta_{\rm g} \psi(x)=&\,-i\,\omega(x)\psi(x)\,\,, \\[3mm]
\delta_{\rm g} \bar{\psi}(x)=&\,i\,\omega(x) \bar{\psi}(x)\,\,, \nonumber
\end{align}
which are broken by the gauge-fixing and infrared regulator terms.

Subjected  to the $U(1)$ gauge transformations (\ref{gaugetransf}), the action
$\Sigma$ (\ref{1}) transforms as given by the following Ward identity:
\begin{equation}
{\mathscr W}_{\rm g}\Sigma= -\frac{1}{e\xi}\,
\bigl(\square+e\xi\lambda^2\bigr)\partial_\mu A^\mu\,\,,
\label{4}
\end{equation}
with the Ward operator associated to the gauge transformations
\begin{equation}
{\mathscr W}_{\rm g}(x)=-\frac{1}{e}\,\partial_{\mu}\frac{\delta}{\delta A_{\mu}(x)}+
i\,\bar{\psi}(x)\frac{\overrightarrow{\delta}}{\delta \bar{\psi}(x)}-
i\,\frac{\overleftarrow{\delta}}{\delta\psi(x)}\psi(x)\,\,.
\label{5}
\end{equation}
Note that the right-hand side of (\ref{4}) being linear in the quantum field
$A_\mu$, will not be submitted to renormalization, {\it i.e.}, it will remain 
a classical breaking \cite{Ref16',piguet}.

On the other hand, the Lorentz symmetry is broken by the presence of the 
constant vector $b_\m$. The fields $A_\m$ and $\p$ transform under
infinitesimal Lorentz transformations $\d x^\m$ =$\e^\m{}_\n x^\n$,
with $\e_{\m\n}=-\e_{\n\m}$, as
\beq\ba{lll}
\delta_\rL A_\mu &=-\e^\l{}_\n x^\n\pa_\l A_\m + \e_\m{}^\n A_\n
&\equiv \frac12 \e^{\a\b} \d_{\rL\a\b} A_\m\,,\\[3mm]
\delta_{\rm L}\p  &=-\e^\l{}_\n x^\n\pa_\l \p
-\frac{i}{4}\e^{\m\n}\s_{\m\n}\p 
&\equiv \frac12 \e^{\a\b} \d_{\rL\a\b} \p\,, 
\ea\eqn{lor-var-A-psi}
where $\sigma_{\mu\nu}=\frac{i}{2}\,[\gamma_\mu,\gamma_\nu]$.

It should be noticed that the Lorentz breaking \equ{SB} 
is not linear in the dynamical fields, therefore will be renormalized.
It is however a ``{\it soft breaking},'' since its UV power-counting
dimension\footnote{The UV power-counting dimensions of $A_\m$ and $\p$
are $d_A=1$ and $d_\p=\frac32$.} is less than 4, namely 3. 
According to Symanzik~\cite{Ref16,Ref16'}, a theory with soft
symmetry breaking is renormalizable if the radiative corrections do not
induce a breakdown of the symmetry by terms of UV power-counting
dimension equal to 4 -- called hard breaking terms. Concretely, according to 
the Weinberg theorem~\cite{weinberg,zimm-clark-low,piguet-rouet}, this
means that the  symmetry of the theory in the asymptotic deep Euclidean
region of momentum space is preserved by the radiative corrections.
In order to control the Lorentz breaking and, in particular, its 
power-counting properties, following Symanzik~\cite{Ref16,Ref16'}, 
and~\cite{schweda} for the specific case of Lorentz breaking, we introduce
an external field $\b_\m(x)$, of dimension 1 and transforming under Lorentz
transformations according to
\beq
\delta_{\rm L}\b_\mu(x)=-\e^\l{}_\n x^\n\pa_\l \b_\m(x)+
\e_\m{}^\n (\b_\n(x)+b_\n)
\equiv \frac12 \e^{\a\b} \d_{\rL\a\b} \b_\m(x) \,.
\eqn{lor-var-beta}
The functional operator which generates these transformations reads
\beq
{\mathscr W}_{\rL\a\b}
=\int d^4x\,\sum_{\vf=A_\m,\p,\bar\p,\b}\d_{\rL\a\b}\vf(x)\frac{\d}{\d \vf(x)}\,.
\eqn{Lor-w-op}
Redefining the action by adding a term in $\b_\mu$:
\beq
\tilde\S = \S - \int d^4x\,\b_\m\bar\p\g_5\g^\m\p\,,
\eqn{25}
one easily checks the classical Ward identity
\beq
{\mathscr W}_{\rm L\a\b}\tilde\Sigma=0\,,
\eqn{tilde-Lor-Ward}
which, at $\b_\m=0$, reduces to the broken Lorentz Ward identity
\beq
{\mathscr W}_{\rm L\a\b}\Sigma = 
\e_\m{}^\n b_\n\int d^4x\,\bar\p\g_5\g^\m\p\,\,.
\eqn{Lor-Ward}
The external field $\beta_\mu(x)$ being coupled to
a gauge invariant expression (the axial current: 
$j_5^\mu=\bar{\psi}\gamma_{5}\gamma^{\mu}\psi$), 
we take it to be gauge invariant in order to preserve gauge invariance,
\be
{\mathscr W}_{\rm g}\int d^4x\,\b_\m\bar\p\g_5\g^\m\p
=0 ~\Longrightarrow~ \delta_{\rm g}\beta_\mu(x)=0\,\,.
\ee
Therefore, it follows that the action $\widetilde{\Sigma}$ (\ref{25}) 
satisfies the same gauge Ward identity 
(\ref{4}) as the action $\Sigma$ (\ref{1}), namely: 
\begin{equation}
{\mathscr W}_{\rm g}\widetilde{\Sigma}=
-\frac{1}{e\xi}\,\bigl(\square+e\xi\lambda^2\bigr)\partial_\mu A^\mu\,\,.
\label{4'}
\end{equation}

\section{Quantization}
\hspace*{\parindent}
In this section, we present the perturbative quantization of the extended QED
theory, using the algebraic renormalization procedure (see~\cite{piguet} for
a review of the method and references to the original literature).
Our aim is to prove that the full quantum theory has the same properties
as the classical theory, {\it i.e.} prove that the Ward identities, associated
to the gauge symmetry \equ{4'} and to the Lorentz symmetry \equ{tilde-Lor-Ward},
are satisfied to all orders of perturbation theory. In order to study the
renormalizability of models characterized by a system of Ward identities,
without referring to any special regularization procedure, two steps
must be followed~\cite{piguet}: In the first step, we compute the possible
anomalies  of the Ward identities through an analysis of the Wess-Zumino
consistency condition. Next, we  check the stability of the classical
action -- which ensures that the quantum corrections do not
produce counterterms corresponding to the renormalization of parameters
not already present in the classical theory.

\subsection{Wess-Zumino Condition: In Search for Anomalies}
\hspace*{\parindent}
The perturbative expansion\footnote{Perturbation theory as usual is ordered
according to the number of loops in the Feynman graphs or, equivalently, to
the powers of $\hbar$.} of the vertex functional:
\be
\Gamma = \sum_{n\geq 0}\hbar \Gamma_n\,\,,
\ee
 is such that it coincides with the classical action in the classical limit:
\be
\Gamma = 
\Gamma_0 + {\mathscr O}(\hbar)\,\,,
\ee
where $\Gamma_0=\widetilde\Sigma$ (\ref{25}).

We have to demonstrate that, at the quantum level, the theory fulfills 
perturbatively, to all orders, the gauge and Lorentz Ward identities:
\be
{\mathscr W}_{\rm g}{\Gamma}=
-\frac{1}{e\xi}\,\bigl(\square+e\xi\lambda^2\bigr)\partial_\mu A^\mu 
\label{quantum-WI-gauge}
\ee 
and
\be
{{\mathscr W}}_{{\rm L}\a\b}{\Gamma}=0\,\,,
\label{quantum-WI-Lorentz}
\ee
-- whose classical counterparts are given by \equ{tilde-Lor-Ward} and \equ{4'} --
together with the normalization conditions:
\beq\ba{ll}
\G_{\ol\psi\psi}(\sl p)\bigg|_{{\ssl p}=m}=0 \,,\quad
&\dpad{}{\sl p}\G_{\ol\psi\psi}(\sl p)
\bigg|_{{\ssl p}=m}=1 \,,\\[3mm]
\dpad{}{p^2}\G_{A^TA^T}(p^2)\bigg|_{p^2=\lambda^2}=1\,,\quad
&-\frac14 {\rm Tr}[\gamma^\mu\gamma^5
\G_{\b_\m\ol\psi\psi}(0,\sl p)]\bigg|_{{\ssl p}=m}=1 \,.
\ea\eqn{normcond}
These four normalization conditions define four of the seven parameters
of the theory, namely the fermion mass $m$ and the amplitudes of the
fields $\p$, $A_\m$ and $\b_\m$. The remaining parameters $e$, $\xi$,
and $\lambda^2$ are defined as the coefficients appearing explicitly in
the Ward identities \equ{quantum-WI-gauge} and \equ{quantum-WI-Lorentz}.

We assume that an ultraviolet subtraction scheme, 
such as the BPHZ \cite{zimm-clark-low,piguet-rouet}, may be applied.
It is well-known that the use of   such a 
subtraction scheme may break the symmetries of the theory -- 
this will  certainly  occur if no invariant regularization procedure is 
available. However, those possible breakings are fully governed by 
the Quantum Action Principle (QAP) \cite{qap,piguet-rouet},  which implies, here: 
\begin{align}
&{\mathscr W}_{\rm g}{\Gamma}+
\frac{1}{e\xi}\,\bigl(\square+e\xi\lambda^2\bigr)\partial_\mu A^\mu 
= \Delta_{\rm g}\cdot\Gamma = \Delta_{\rm g} + {\mathscr O}(\hbar)\,\,,
\label{QAP1}\\[3mm] 
&{{\mathscr W}}_{{\rm L}\a\b}{\Gamma}=\Delta_{{\rm L}\a\b}\cdot\Gamma 
= \Delta_{{\rm L}\a\b} + {\mathscr O}(\hbar)\,\,,\label{QAP2}
\end{align}
where $\Delta_{\rm g}(x)$ and $\Delta_{{\rm L}\a\b}(x)$ are 
local insertions with their UV dimensions bounded by $d_{\Delta_{\rm g}}\leq4$
and $d_{\Delta_{{\rm L}\a\b}}\leq4$, respectively. 
The Ward identity operators ${\mathscr W}_{\rm g}$ and 
${{\mathscr W}}_{\rm L\alpha\beta}$ obey the following commutation rules:
\begin{align}
&[{\mathscr W}_{\rm g}(x),{\mathscr W}_{\rm g}(y)]=0\label{WgWg}\,,\\[3mm]
&[{\mathscr W}_{\rm g}(x), 
{{\mathscr W}}_{\rm L\a\b}(y)]= 0\,,\label{WgWL} \\[3mm]
&[{{\mathscr W}}_{\rm L\alpha\beta}(x),
{{\mathscr W}}_{\rm L\gamma\delta}(y)]=
\Bigl\{\eta_{\alpha\delta}{{\mathscr W}}_{\rm L\beta\gamma}(x)+
\eta_{\beta\gamma}{{\mathscr W}}_{\rm L\alpha\delta}(x)-
\eta_{\alpha\gamma}{{\mathscr W}}_{\rm L\beta\delta}(x)- 
\eta_{\beta\delta}{{\mathscr W}}_{\rm L\alpha\gamma}(x)\Bigr\}\delta(x-y)\,,
\label{WLWL}
\end{align}
where the latter commutation relation  is that of the Lorentz algebra. 
By adopting the notation 
$[\a\b]=i$ ($i=1,...,6$) for any (antisymmetric) pair of Lorentz indices of 
the Ward operator ${{\mathscr W}}_{\rm L\alpha\beta}$, the eq.(\ref{WLWL}) 
can be rewritten as
\be
[{{\mathscr W}}_{{\rm L} i}(x),{{\mathscr W}}_{{\rm L} j}(y)]=
f^k_{ij}{{\mathscr W}}_{{\rm L} k}\delta(x-y)\,\,.
\label{WLWLij}
\ee 
Now, through (\ref{WgWg}), (\ref{WgWL}) and (\ref{WLWL}), the insertions 
$\Delta_{\rm g}$ and $\Delta_{{\rm L}i}$ appearing in (\ref{QAP1}) and (\ref{QAP2}),
which are local field polynomials, satisfy the following Wess-Zumino consistency
conditions: 
\be\ba{l}
{\mathscr W}_{\rm g}(x)\Delta_{\rm g}(y)
-{\mathscr W}_{\rm g}(y)\Delta_{\rm g}(x)=0\,,\\[3mm]
{\mathscr W}_{\rm g}(x)\Delta_{{\rm L}j}(y)
-{\mathscr W}_{{\rm L}j}(y)\Delta_{\rm g}(x)=0\,,\\[3mm]
{\mathscr W}_{{\rm L}i}(x)\Delta_{{\rm L}j}(y)-
{\mathscr W}_{{\rm L}j}(y)\Delta_{{\rm L}i}(x)
=f^k_{ij}\Delta_{{\rm L}k}(x)\delta(x-y)\,\,.
\ea\label{consistency}\ee
In the present case, it turns out to be convenient to proceed step by step,
beginning with the Lorentz Ward identity. Its validity has been proved
in~\cite{schweda}, using Whitehead's lemma for semi-simple Lie groups,
shown {\it e.g.} in~\cite{brs-break}, which states the vanishing of the
first cohomology of such groups. In our context, this means that the general
solution of the last of equations \equ{consistency} has the form
\be
\Delta_{{\rm L}i}(x)={\mathscr W}_{{\rm L}i}(x)\widehat\Delta_{\rm L}\,\,, 
\label{DeltaL}\ee
with  $\widehat\Delta_{\rm L}$ an integrated local insertion of UV dimension
bounded by $d_{\widehat\D_{\rm L}}\le 4$. $\widehat\Delta_{\rm L}$ can then be
reabsorbed in the action as a -- noninvariant -- counterterm, order by order,
thus establishing the Lorentz Ward identity \equ{quantum-WI-Lorentz}
perturbatively at each order.

Let us now turn to the gauge Ward identity \equ{quantum-WI-gauge}.
Since we can now assume the validity of the Lorentz Ward identity, the
consistency equations \equ{consistency} reduce to
\be\ba{l}
{\mathscr W}_{\rm g}(x)\Delta_{\rm g}(y)
-{\mathscr W}_{\rm g}(y)\Delta_{\rm g}(x)=0\,,\\[3mm]
{\mathscr W}_{{\rm L}j}(x)\Delta_{\rm g}(y)=0\,.
\ea\label{consistency-gauge}\ee
The general solution is well-known~\cite{brs-gauge,piguet}: it is the
(Abelian) Adler-Bardeen-Bell-Jackiw an\-om\-aly~\cite{abbj} -- up to
terms which are gauge variations of integrated local insertions $\hat\Delta$
which can be reabsorbed as counterterms:
\be
\Delta_{\rm g}(x) = {\mathscr W}_{\rm g}(x)\hat\Delta +
r\,\e_{\m\n\rho\lambda}F^{\m\n}F^{\rho\lambda}(x)\,.
\ee
The anomaly coefficient $r$ being not renormalized~\cite{adl-bard,piguet},
{\it i.e.}, it is zero if it vanishes at the 1 loop order, it suffices to check
its vanishing at this order. But this is obvious, since  the potentially
dangerous axial current $j_5^\mu=\bar{\psi}\gamma_{5}\gamma^{\mu}\psi$ is
only coupled to the external field $\b_\mu$ -- and not to any quantum field
of the theory, which means that in fact no gauge anomaly can be
produced~\cite{piguet-rouet,Ref11,Ref11'}. Thus, the gauge Ward identity is
preserved at the quantum level. 

\subsection{Stability: In Search for Counterterms}
\hspace*{\parindent}
For the quantum theory the stability corresponds to the fact that the radiative 
corrections --  the Ward identities being supposed to hold at this stage --
can be reabsorbed by a redefinition of the initial parameters of the theory. 
As it is well known~\cite{piguet}, it suffices to check the stability of
the invariant classical action. In order to do so, one perturbs 
the action $\widetilde{\Sigma}$ \equ{25} 
by an arbitrary integrated local $\widetilde{\Sigma}^{\rm c}$: 
\begin{equation}
\widehat{\Sigma} = \widetilde{\Sigma} +\epsilon\,\widetilde{\Sigma}^{\rm c}\,\,,
\label{stabe}
\end{equation}
where $\epsilon $ is an infinitesimal parameter and
the functional $\widetilde{\Sigma}^{\rm c}$ has the same quantum numbers (dimension,
discrete symmetries) as 
the classical action. One then requires the deformed action $\widehat{\Sigma}$ 
to obey all the classical Ward identities:
\begin{equation}
{\mathscr W}_{\rm g}(\widetilde{\Sigma}+\epsilon \,\widetilde{\Sigma}^{\rm c})
={\mathscr W}_{\rm g}(\widetilde{\Sigma})+\epsilon\,
{\mathscr W}_{\rm g}\widetilde{\Sigma}^{\rm c}=
-\frac{1}{e\xi}\,\bigl(\square+e\xi\lambda^2\bigr)\partial_\mu A^\mu\,\,,
\label{acaope}
\end{equation}
and 
\begin{equation}
{{\mathscr W}}_{{\rm L}\a\b}(\widetilde{\Sigma}+
\epsilon\,\widetilde{\Sigma}^{\rm c})
={{\mathscr W}}_{{\rm L}\a\b}(\widetilde{\Sigma})+\epsilon\,
{{\mathscr W}}_{{\rm L}\a\b}\widetilde{\Sigma}^{\rm c}=0\,\,.
\label{acaope'}
\end{equation}
Then $\widetilde{\Sigma}^{\rm c}$ is subjected to the following set of
constraints:
\begin{equation}
{\mathscr W}_{\rm g}\widetilde{\Sigma}^{\rm c}=0\,\,,\quad
{{\mathscr W}}_{{\rm L}\a\b}\widetilde{\Sigma}^{\rm c}=0\,\,,
\label{acaope1}
\end{equation}
therefore, the counterterm $\widetilde{\Sigma}^{\rm c}$ must be
{\it symmetric} under the gauge and Lorentz symmetries as shown 
by eqs.(\ref{acaope1}), as well as invariant under charge conjugation.

The most general Lorentz invariant counterterm $\widetilde{\Sigma}^{\rm c}$,
{\it i.e.}, the most general Lorentz invariant field polynomial with UV dimension 
bounded by $d\leq4$ is given by an arbitrary superposition of the following 
-- {\it integrated} -- monomials:  
\begin{align*}
&\bigg\{ \bar{\psi}\psi\,\,,\,\,
\bar{\psi}\gamma^{\mu}\partial_{\mu}\psi\,\,,\,\,
\bar{\psi}\gamma^{\mu}A_{\mu}\psi\,\,,\,\,
(\beta_{\mu}(x)+b_\mu)\bar{\psi}\gamma_{5}\gamma^{\mu}\psi\,\,,\,\,
\partial_{\mu}A_{\nu}\partial^{\mu}A^{\nu}\,\,,\,\,
\partial_{\mu}A_{\nu}\partial^{\nu}A^{\mu}\,\,,\,\, \\[3mm]
&\qquad A_{\mu}A^{\mu}\,\,,\,\,
A_{\mu}A^{\mu}A_{\nu}A^{\nu}\,\,,\,\,
(\beta_{\mu}(x)+b_\mu)A^{\mu}\partial_{\nu}A^{\nu}\,\,,\,\,
\,\,(\beta_{\mu}(x)+b_\mu)A^{\nu}\partial_{\nu}A^{\mu}\,\,,\\[3mm]
&\qquad(\beta_{\mu}(x)+b_\mu)A_{\nu}\partial^{\mu}A^{\nu}\,\,,\,\,
((\beta_{\mu}(x)+b_\mu)A^{\mu})^{2}\,\,,\,\,
\epsilon_{\mu\nu\alpha\beta}(\beta^{\mu}(x)+b^\mu)
A^{\nu}\partial^{\alpha}A^{\beta} \bigg\}\,\,.
\end{align*}
Moreover, {\it gauge invariance} -- represented by the first of Ward identities
\equ{acaope1} -- and the invariance under {\it charge conjugation} select
the following four field polynomials:
\begin{align*}
{\mathscr P}_1=i\bar{\psi}\gamma ^\mu (\partial_{\mu}+ieA_{\mu}) \psi\,\,,\,
{\mathscr P}_2=\bar{\psi}\psi\,\,,\,
{\mathscr P}_3=(\beta_{\mu}(x)+b_\mu)\bar{\psi}\gamma_{5}\gamma^{\mu}\psi\,\,,\,
{\mathscr P}_4=F^{\mu\nu}F_{\mu\nu}\,\,,
\end{align*}
Now, it should be pointed out that, taking into account {\it only} the gauge
Ward identity, ${{\mathscr W}}_{\rm g}\widetilde{\Sigma}^{\rm c}=0$, 
and the charge conjugation invariance, a Chern-Simons-like term,  
$\int d^4x~\epsilon_{\mu\nu\alpha\beta}b^{\mu}A^{\nu}\partial^{\alpha}A^{\beta}$, 
could appear as a possible counterterm to the extended QED (\ref{25}). It is
softly broken Lorentz invariance as expressed by the second of Ward identities
\equ{acaope1}, which rules out this term. More precisely, it is a consequence
of the postulated gauge invariance of the external field $\beta_\m(x)$ 
introduced to characterize the softly broken Lorentz invariance.

Finally, the most general integrated local functional, $\widetilde{\Sigma}^{\rm c}$, 
satisfying the conditions of gauge and Lorentz invariances (\ref{acaope1}), 
and invariant under charge conjugation, is given by:
\begin{eqnarray}
\widetilde{\Sigma}^{\rm c}=\int d^4x\,\sum\limits_{i=1}^4 a_i{\mathscr P}_i(x)
\,\,, \label{poly1}
\end{eqnarray}
where $a_1,...,a_4$ are arbitrary coefficients, fixed by the four normalization 
conditions (\ref{normcond}) order by order in perturbation theory.

\section{Conclusions}
\hspace*{\parindent}
We proved through the use of the algebraic method of renormalization, which
{\it is independent of any kind of regularization scheme}, that
in the case of extended QED (\ref{1}), under the hypothesis discussed in the next
paragraph, a CPT-odd and Lorentz violating Chern-Simons-like term is definitively
{\it not} generated by the radiative corrections. Therefore, if 
the Chern-Simons-like term is absent, from the beginning, at the classical level, 
it will be absent at the quantum level. This result has been obtained through a
careful analysis of the consequences of the symmetries -- Lorentz and gauge
invariance -- taken together in a consistent way.

As we have said in the introduction, various and apparently contradictory claims 
are found in the literature. We must stress that our result is linked to an
assumption we have made, namely that the external vector field
$\b_\m(x)$, introduced in order to control the Lorentz breaking, {\it is gauge
invariant}. Equivalently, our hypothesis has been  that  the axial current to
which $\b_\m(x)$ is coupled and which charaterizes the Lorentz breaking, 
is gauge invariant. As discussed in details in~\cite{Ref2} with the help of
explicit one-loop computations, this choice naturally forbids a Chern-Simons
like counterterm -- whose integrant is not gauge invariant -- if this term is
not already present in the tree aproximation. Our work confirms this point,
unambiguously, to {\it all orders of perturbation theory}. Note that, still according
to~\cite{Ref2}, relaxing the assumption of gauge invariance of
the local axial current and only requiring the invariance of its
spacetime integral, would allow such a counterterm -- and even fix it in
a so-called ``nonperturbative in $b_\n$'' treatment as the authors 
of~\cite{Ref2} show in the one-loop approximation.

As a final remark, it should be noted that the same vanishing result has been proven to all
orders in Ref.~\cite{Ref11}, with similar methods but {\it no explicit use} of the Lorentz
Ward identities. There it is argued that, if the theory is correctly defined through Ward
identities and normalization conditions, no Chern-Simons-like term appears,
without any ambiguity. This is related to the fact that such term, bilinear
in the gauge field, appears in fact as a {\it minor modification} to the
gauge-fixing term. Then, as part of the ``gauge term,'' it is not renormalized. However,
in our opinion, this argument must be better understood, since 
the analysis contained in Ref.~\cite{Helayel} (using the method of spin projectors) shows
that the Chern-Simons-like term is linked to the sector of spin 1, which is the sector
that carries the physical degrees of freedom of the model. This apparently indicates that the
Chern-Simons-like term could not be seen as a minor modification to the term of
gauge-fixing~\cite{Helayel1}.

\subsection*{Acknowledgements}
\hspace*{\parindent}
O.M.D.C. dedicates this work to his kids, Vittoria and Enzo, to his mother,
Victoria, and to Clarisse Czertok. J.M.F. is supported by the Funda\c c\~ao
de Amparo \`a Pesquisa do Estado de Minas Gerais (FAPEMIG) agency. D.H.T.F.
dedicates this work to the Prof. Lincoln Almir Amarante Ribeiro (in memoriam). 
O.P. was supported   in part by the Conselho Nacional de Desenvolvimento
Cient\'{\i}fico e Tecnol\'{o}gico -- CNPq (Brazil) and by the PRONEX project
${\rm N}.^\circ$ 35885149/2006 from FAPES -- CNPq (Brazil).



\begin{thebibliography}{99}

\bibitem{Ref1} S. Coleman and S.L. Glashow, ``{\em High-energy tests of
Lorentz invariance,}'' Phys. Rev. {\bf D59}  (1999) 11608.

\bibitem{Ref2} R. Jackiw and V.A. Kosteleck\'{y}, ``{\em Radiatively Induced Lorentz and CPT Violation in Electrodynamics,}'' Phys. Rev. Lett. {\bf 82}  (1999) 3572.

\bibitem{Ref3} J.-M. Chung, ``{\em Lorentz and CPT-violating Chern-Simons term in the functional integral formalism,}'' Phys. Rev. {\bf D60}  (1999) 127901.

\bibitem{Ref4} J.-M. Chung, ``{\em Radiatively-induced Lorentz and CPT Violating Chern-Simons
term in QED,}'' Phys. Lett. {\bf B461}  (1999) 138.

\bibitem{Ref5} M. P\'{e}rez-Victoria, ``{\em Exact Calculation of the Radiatively Induced Lorentz and CPT Violation in QED,}'' Phys. Rev. Lett. {\bf 83} (1999)  2518.

\bibitem{Ref7} W.F. Chen, ``{\em Understanding radiatively induced Lorentz-CPT violation in differential regularization,}'' Phys. Rev. {\bf D60} (1999)  085007.

\bibitem{Ref8} J.-M. Chung and Phillial Oh, ``{\em Lorentz and CPT-violating Chern-Simons term in the derivative expansion of QED,}'' Phys. Rev. {\bf D60}  (1999) 067702.

\bibitem{Ref9} J.-M. Chung and B. K. Chung, ``{\em Induced Lorentz- and CPT-violating Chern-Simons term in QED: Fock-Schwinger proper time method,}'' Phys. Rev. {\bf D63}
 (2001) 105015.

\bibitem{Ref10} C. Adam and F.R. Klinkhamer, ``{\em Causality and radiatively induced CPT violation,}'' Nucl. Phys. {\bf B513} (2001)  245.

\bibitem{Ref11} G. Bonneau, ``{\em Regularisation: many recipes, but a unique
 principle: Ward identities and normalisation conditions. The case of CPT violation in QED,}''
Nucl. Phys. {\bf B593}  (2001) 398.

\bibitem{Ref11'} G. Bonneau, ``{\em Extended QED with CPT violation: Clarifying some controversies}'' Nucl. Phys. {\bf B764}  (2007) 83.

\bibitem{Ref12} M. P\'{e}rez-Victoria, ``{\em Physical (ir)relevance of ambiguities to Lorentz and CPT violation in QED,}'' JHEP {\bf 04}  (2001)  032.

\bibitem{Ref12'} W.F. Chen, ``{\em Issues on radiatively induced Lorentz and CPT violation in quantum electrodynamics,}'' hep-th/0106035.

\bibitem{Ref12''} Y.A. Sitenko and K. Yu. Rulik, ``{\em On the effective Lagrangian in spinor
electrodynamics with added violation of Lorentz and CPT symmetries,}''
Eur. Phys. J. {\bf C28}  (2003) 405.

\bibitem{Ref13} B. Altschul, ``{\em Gauge invariance and the Pauli-Villars regulator in Lorentz-and CPT-violating electrodynamics,}'' Phys. Rev. {\bf D70}  (2004) 101701.

\bibitem{Ref14} A.P. Ba\^{e}ta Scarpelli, Marcos Sampaio, M.C. Nemes and B. Hiller, 
``{\em Gauge invariance and the CPT and Lorentz induced Chern-Simons-like term 
in extended QED,}'' Eur. Phys. J. {\bf C56}  (2008) 571.

\bibitem{Brito} F.A. Brito, J.R. Nascimento, E. Passos and A.Yu. Petrov, ``{\em The ambiguity-free four-dimensional Lorentz-breaking Chern-Simons action,}'' Phys. Lett. {\bf B664}  (2008) 112.

\bibitem{Ref15} J. Alfaro, A.A. Andrianov, M. Cambiaso, P. Giacconi and R. Soldati,
``{\em Induced Lorentz \& CPT invariance violations in QED,}'' arXiv:0904.3557 [hep-th].

\bibitem{Ref16} K. Symanzik, ``{\em Renormalizable models with simple symmetry breaking: 
1. Symmetry breaking by a source term,}'' Commun. Math. Phys. {\bf 16} 48 (1970). 

\bibitem{Ref16'} K. Symanzik, in Carg\`ese Lectures in Physics (1970). vol. 5, ed. D. Bessis
(Gordon \& Breach. 1972).

\bibitem{brs-break} C. Becchi, A. Rouet and R. Stora, 
``{\em Renormalizable Theories with Symmetry Breaking,}'' 
in Field Theory, Quantization and Statistical Physics. In memory of Bernard Jouvel,
ed. E. Tirapegui, D. Reidel Publishing Co.; \\
C. Becchi, A. Rouet and R. Stora, 
``{\em Renormalizable Models with Broken Symmetries}'',
in ``Renormalization Theory'', eds. G. Velo and A.S. Wightman, 
D. Reidel Publishing Co., 1976.

\bibitem{piguet} O. Piguet and S.P. Sorella, ``{\em Algebraic Renormalization,}'' 
Lecture Notes in Physics, m28, Springer-Verlag, 1995.

\bibitem{greenberg}  O.W. Greenberg, ``{\em CPT violation implies violation of Lorentz invariance,}'' Phys. Rev. Lett. {\bf 89}  (2002) 231602. 

\bibitem{piguet-rouet} O. Piguet and A. Rouet, ``{\em Symmetries in perturbative quantum field
 theory,}'' \prep{76}{81}{1}

\bibitem{low-schroer} J.H. Lowenstein and B. Schroer,``{\em Gauge invariance and
Ward identities in a massive vector meson model,}'' \pr{D6}{72}{1553} and
``{\em Comment on the absence of radiative corrections to the anomaly of the axial-vector currentl,}''  \pr{D7}{73}{1929}

\bibitem{itzykson} C. Itzykson and J.-B. Zuber, ``{\em Quantum Field Theory,}'' Physics Series, 
McGraw-Hill, 1980. 

\bibitem{weinberg} S. Weinberg, ``{\em High-energy behavior in quantum field
theory,}''  \pr{118}{60}{838}

\bibitem{zimm-clark-low} W. Zimmermann, ``{\em Convergence of Bogolyubov's method of renormalization in momentum space,}''  \cmp{15}{69}{208}\\
J.H. Lowenstein, ``{\em Convergence Theorems for Renormalized Feynman Integrals
with Zero-Mass Propagators,}''  \cmp{47}{76}{53}\\
 T.E. Clark and J.H. Lowenstein,  ``{\em Generalization of Zimmermann's Normal-Product
Identity,}'' \np{B113}{76}{109}

\bibitem{schweda}  H. Balasin, M. Schweda, M. Stierle  and O. Piguet,
``{\em The Cohomology Problems Of Rigid Lorentz Transformations 
in Axial Gauge Theories}'', \pl{B215}{88}{328}

\bibitem{qap} J.H. Lowenstein, ``{\em Normal product quantization of currents in
Lagrangian field theory,}'' \pr{D4}{71}{2281} and ``{\em Differential vertex operations
in Lagrangian field theory}'' \cmp{24}{71}{1}\\
Y.M.P. Lam, ``{\em Perturbation Lagrangian theory for scalar fields: Ward-Takahasi identity
and current algebra,}'' \pr{D6}{72}{2145} and ``{\em Equivalence theorem on
Bogolyubov-Parasiuk-Hepp-Zimmermann renormalized Lagrangian field theories,}''
\pr{D7}{73}{2943}\\
T.E. Clark and J.H. Lowenstein,  ``{\em Generalization of Zimmermann's Normal-Product
Identity,}'' \np{B113}{76}{109}

\bibitem{brs-gauge}  C. Becchi, A. Rouet and R. Stora, ``{\em Renormalization of Gauge Theories,}''
\annp{98}{76}{287} 

\bibitem{abbj} S. L. Adler, ``{\em Axial vector vertex in spinor electrodynamics,}''
\pr{117}{69}{2426} \\
  J. S. Bell and R. Jackiw, ``{\em A PCAC puzzle: $\pi_0 \rightarrow$ gamma gamma in the sigma
model,}''  \nc{60}{69}{47} \\
  J. Schwinger, ``{\em On gauge invariance and vacuum polarization,}''
\pr{82}{51}{664}

\bibitem{adl-bard} S.L. Adler and W.A. Bardeen, ``{\em Absence of higher order corrections
in the anomalous axial vector divergence equation,}''  \pr{182}{69}{1517} \\
W. A. Bardeen,   ``{\em Renormalization of Yang-Mills fields and application to particle
 physics,}'' {\em C.N.R.S. (Marseille) report 72/p470, 29 (1972)}.\\
 W. Bardeen, {\em proceedings of the 16th International Conference on High
Energy Physics, Fermilab 1972, vol. II, p.295}.\\
A. Zee, ``{\em Axial Vector Anomalies And The Scaling Property Of Field Theory,}''
\prl{29}{72}{1198}\\
A. Blasi, O. Piguet and S. P. Sorella, ``{\em Landau gauge and finiteness,}''
\np{B356}{91}{154}

\bibitem{Helayel}  A. P. Baeta Scarpelli, H. Belich, J. L. Boldo and J. A. Helay\"el-Neto,
``{\em Aspects of Causality and Unitarity and Comments on Vortexlike Configurations in an
Abelian Model with a Lorentz Breaking Term,}'' Phys. Rev, {\bf D67} (2003) 085021.

\bibitem{Helayel1} J. A. Helay\"el-Neto, private communications.

\end{thebibliography}
\end{document}